\documentclass{svmult}

\usepackage{graphicx}
\usepackage{graphics}
\usepackage{makeidx}
\usepackage{multicol}
\usepackage{footmisc}

\usepackage[centertags]{amsmath}
\usepackage{amsfonts}
\usepackage{amssymb}
\usepackage{url}
\usepackage{subfigure}

\begin{document}

\title*{Computation with competing patterns in Life-like automaton}


\author{Genaro J. Mart{\'i}nez\inst{1,2} \and Andrew Adamatzky\inst{2} \and Kenichi Morita\inst{3} \and Maurice Margenstern\inst{4}}
 
\institute{Centro de Ciencias de la Complejidad, Universidad Nacional Aut\'onoma de M\'exico, M\'exico DF. \\
\url{genaro.martinez@uwe.ac.uk}
\and
Bristol Institute of Technology, Bristol, United Kingdom.
\url{andrew.adamatzky@uwe.ac.uk}
\and
Hiroshima University, Higashi-Hiroshima 739-8527, Japan.
\url{morita@iec.hiroshima-u.ac.jp}
\and
Laboratoire d'Informatique Th\'eorique et Appliqu\'ee, 
Universit\'e de Metz, Metz Cedex, France. \\
\url{margens@univ-metz.fr}
}

\authorrunning{Mart{\'i}nez G. J. et al.}

\maketitle

\begin{abstract}
We study a Life-like cellular automaton rule $B2/S2345$ where a cell in state `0' takes state `1' if it has exactly two neighbors in state `1' and the cell remains in the state `1' if it has between two and five neighbors in state `1.' This automaton is a discrete analog spatially extended chemical media, combining both properties of sub-excitable and precipitating chemical media. When started from random initial configuration $B2/S2345$ automaton exhibits chaotic behavior. Configurations with low density of state `1' show emergence of localized propagating patterns and stationary localizations. We construct basic logical gates and elementary arithmetical circuits by simulating logical signals with mobile localizations reaction propagating geometrically restricted by stationary non-destructible localizations. Values of Boolean variables are encoded into two types of patterns --- symmetric ({\sc False}) and asymmetric ({\sc True}) patterns -- which compete for the `empty' space when propagate in the channels. Implementations of logical gates and binary adders are illustrated explicitly.
\end{abstract}

\section{Introduction}
\label{introduction}

Computational universality of Conway's Game of Life (GoL) cellular automaton~\cite{kn:Gard70} has been already demonstrated by various implementations. Most famous designs are realization of functionally complete set of logical functions~\cite{kn:Ren03}, register machine~\cite{kn:BCG82}, direct simulation of Turing machine~\cite{kn:Chap02, kn:Ren02}, and design of a universal constructor~\cite{kn:Gou09}. These implementations use principles of collision-based computing~\cite{kn:BCG82, kn:Ada02} where information is transferred by localizations (gliders) propagating in an architecture-less, or `free,' space. The theoretical results regarding GoL universality is only the first half-step in a long journey towards real-world implementation of the collision-based computers as unconventional computing \cite{kn:Toff98, kn:Ada01}. Controllability of signals is the first obstacle to overcome. Despite their mind-whirling elegance and complexity-wise efficiency of implementation the `free-space' computing circuits are difficult to fabricate in physical or chemical materials~\cite{kn:ACA05} because propagating localizations (solitons, breathers, kinks, wave-fragments) are notoriously difficult to manipulate, maintain and navigate. 

The easiest way to control patterns propagating in a non-linear medium circuits is to constrain them geometrically. Constraining the media geometrically is a common technique used when designing computational schemes in spatially extended non-linear media. For example `strips' or `channels' are constructed within the medium (e.g. excitable medium) and connected together, typically using arrangements such as $T$-junctions. Fronts of propagating phase (excitation) or diffusive waves represent signals, or values of logical variables. When fronts interact at the junctions some fronts annihilate or new fronts emerge. The propagations in the output channels represent results of the computation. 

The geometrical-constraining approach is far from being graceful (particularly, comparing to collision-based paradigm~\cite{kn:Ada02, kn:IM00, kn:MMI99}) but simple and practical enough to be used by experimental scientists and engineers. The geometrical constraining is successfully applied in design of several laboratory prototypes of chemical (all based on Belousov-Zhabotinsky reaction) computing devices: logical gates~\cite{kn:TS95, kn:SG01}, diodes~\cite{kn:KYA97, kn:DAK98, kn:MYI01}, counters~\cite{kn:GYI03}, coincidence detectors~\cite{kn:GG03}, and memory~\cite{kn:MYI01}.

What members of GoL family offer us most `realistic' approximation of dynamical processes in spatially extended chemical computers? In its original form,  GoL automaton is a discrete analog of sub-excitable chemical media~\cite{kn:BE03, kn:KPK90}. Localized wave-fragments (reaction-diffusion dissipative solitons) in sub-excitable Belousov-Zhabotinsky reaction~\cite{kn:BE03, kn:KPK90} are represented by gliders~\cite{kn:Gard70, kn:Wain73, kn:GM96, kn:MLH97, kn:PW85, kn:AMS05} in GoL cellular automaton. See examples of direct comparisons between experimental chemical laboratory results and cellular automaton models in \cite{kn:CTS09}.

There is also a family of Life-life rules, where `cells never die,' the state `1' is an absorbing state.  This is the family of \emph{Life without Death} (LwD), invented by Griffeath and Moore in~\cite{kn:GM96}. In the LwD automaton we  observe propagating patterns, formed due to rule-based restrictions on propagation similar to that in sub-excitable chemical media and plasmodium of \emph{Physarym polycephalum}~\cite{kn:Ada07}, but no complicated periodic structures or global chaotic behavior occurs. The LwD family of cell-state transition rules is an automaton equivalent of the precipitating chemical systems. This is demonstrated in our phenomenological studies of semi-totalistic and precipitating CA~\cite{kn:AMS05}, where we selected a set of rules Life $2c22$, identified by periodic structures~\cite{kn:MMZ05}. The clans closest to the family $2c22$ are \emph{Diffusion Rule} (Life rule $B2/S7$)~\cite{kn:MAM}, all they also into of a big cluster named as Life $dc22$.\footnote{\url{http://uncomp.uwe.ac.uk/genaro/Life_dc22.html}}

In present chapter we study a Life-like cellular automaton, which somewhat imitates properties of both excitable and precipitating reaction-diffusion chemical systems, and who how to implement a sensible computation in such type of cellular automata. We develop a model where the channels are constructed using static patterns and computation is implemented by propagating patterns of precipitation. The results are based on our previous studies of reaction-diffusion analogs $B2/S23456$ \cite{kn:MAC08} and $B2/S2345678$ \cite{kn:MAM08}.

\section{Life rule $B2/S2345$}
\label{activity}

Life rule $B2/S2345$ is described as follows. Each cell takes two states `0' (`dead') and `1' (`alive'), and updates its state depending on its eight closest neighbors:

\begin{enumerate}
\item Birth: a central cell in state 0 at time step $t$ takes state 1 at time step $t+1$ if it has exactly two neighbors in state 1.
\item Survival: a central cell in state 1 at time $t$ remains in the state 1 at time $t+1$ if it has two, three, four or five live neighbors.
\item Death: all other local situations.
\end{enumerate}

Once a resting lattice is perturbed (few cells are assigned live states), patterns of states 1 emerge, grow and propagate on the lattice quickly.

A general behavior of rule $B2/S2345$ can be well described by a mean field polynomial and its fixed points. Mean field theory is a proved technique for discovering statistical properties of CA without analyzing evolution spaces of individual rules~\cite{kn:GV87, kn:Mc90, kn:CM91}. The method assumes that elements of the set of states are independent, uncorrelated between each other in the rule's evolution space. Therefore we can study probabilities of states in neighborhood in terms of probability of a single state (the state in which the neighborhood evolves), thus probability of a neighborhood is the product of the probabilities of each cell in the neighborhood.

The mean field polynomial for rule $B2/S2345$ is as follows: 

\begin{equation}
p_{t+1} = 28p_t^3q_t^6+56p_t^4q_t^5+56p_t^6q_t^3+70p_t^5q_t^4+28p_t^2q_t^7.
\label{meanF}
\end{equation}

\begin{figure}[th]
\centering
\includegraphics[width=0.72\textwidth]{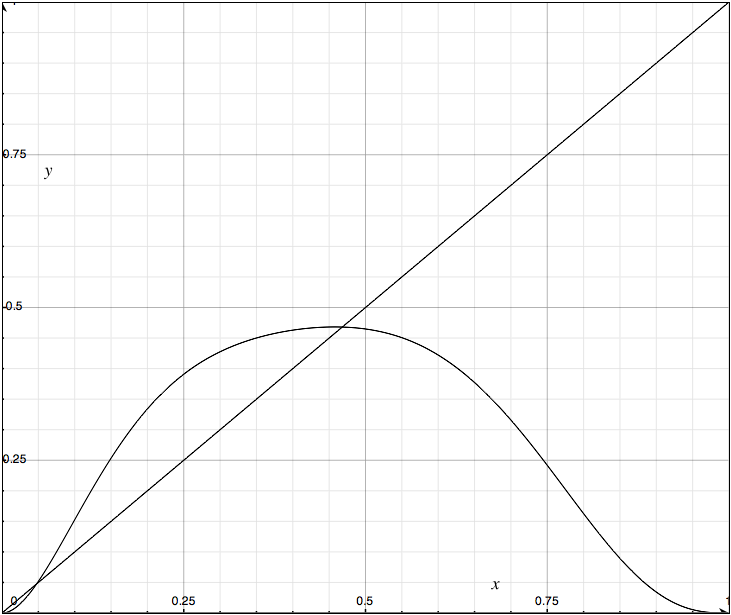}
\caption{Mean field curve for $B2/S2345$.}
\label{meanF-1}
\end{figure}

\begin{figure}[th]
\begin{center}
\subfigure[]{\scalebox{0.359}{\includegraphics{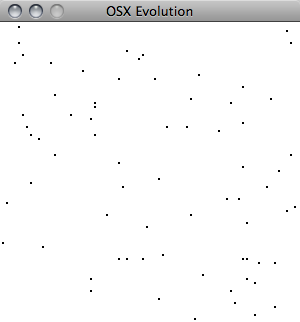}}} 
\subfigure[]{\scalebox{0.359}{\includegraphics{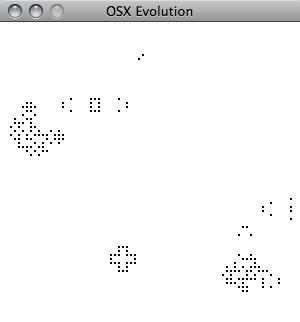}}} 
\subfigure[]{\scalebox{0.359}{\includegraphics{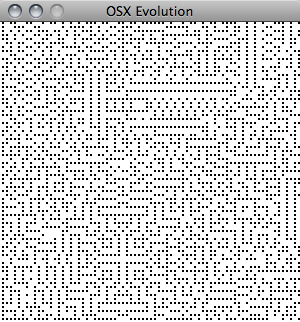}}}
\end{center}
\caption{Snapshots of automaton configurations of $300 \times 300$ cells showing activity of particles in $B2/S2345$. (a)~Initial random configuration, density of state `1' is 1.2\%. (b)~Just after eight generations gliders and nucleation patterns emerge. (c)~540th generations.}
\label{meanF-2}
\end{figure}

High densities of regions dominated by state 1 correspond to maximum point near $p=0.35$. The average density is reached with one stable fixed point $p=0.46$ when automaton find its stability around of 37\% of cells in state `1.' Some interesting behavior can be found in extreme unstable fixed point when $p=0.04$ (complex class, see mean field classification in \cite{kn:Mc90}). Looking on configurations with less than 4\% of cells in state `1' we can observer gliders, oscillators, and still life patterns. Thus unstable fixed points (Fig.~\ref{meanF-1}) represent evidence of complex behavior as mobile localizations, gliders and small oscillators emerging in the automaton development. Figure~\ref{meanF-2} demonstrates a typical evolution of $B2/S2345$ starting with very small initial densities of state `1.' Mobile localizations emerge but do not survive for a long time.

\begin{figure}[th]
\begin{center}
\subfigure[]{\scalebox{0.25}{\includegraphics{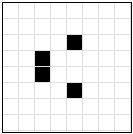}}} \hspace{0.7cm}
\subfigure[]{\scalebox{0.25}{\includegraphics{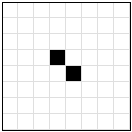}}} \hspace{0.7cm}
\subfigure[]{\scalebox{0.25}{\includegraphics{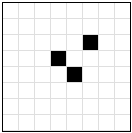}}} \hspace{0.7cm}
\subfigure[]{\scalebox{0.25}{\includegraphics{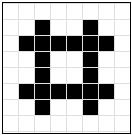}}}
\end{center}
\caption{Basic periodic structures in $B2/S2345$: (a)~glider,  (b)~flip-flop oscillator, (c)~blinker oscillator, and (d)~still life configuration.}
\label{basicPatterns}
\end{figure}

A set of minimal particles, or basic periodic structures, in rule $B2/S2345$ include one glider, two oscillators (one blinker and one flip-flop configurations), and one still life configuration (see Fig.~\ref{basicPatterns}). The still life patterns \cite{kn:Mc88, kn:Cook03} represent precipitation of an abstract reaction-diffusion chemical system imitated by rule $B2/S2345$ automaton. The still life blocks are not affected by their environment however they do affect their environment~\cite{kn:MAC08,kn:MAM08}). Therefore the still life patterns can be used to build channels, or wires, for signal propagation. 

\subsection{Indestructible pattern in $B2/S2345$}
\label{stillLifep}

Some patterns amongst still life patterns in the rule $B2/S2345$ belong to a class of \emph{indestructible patterns} (sometimes referred to as `glider-proof' patterns) which cannot be destroyed by any perturbation, including collisions with gliders. A minimal indestructible pattern, still life occupying a square of $6 \times 6$ cells, is shown in Fig.~\ref{basicPatterns}d.  

\begin{figure}
\begin{center}
\subfigure[]{\scalebox{0.23}{\includegraphics{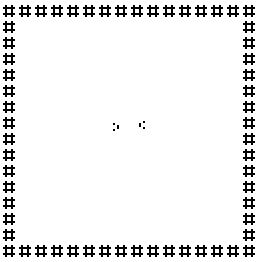}}} \hspace{1.0cm}
\subfigure[]{\scalebox{0.23}{\includegraphics{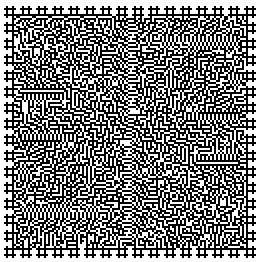}}} \hspace{1.0cm}
\subfigure[]{\scalebox{0.23}{\includegraphics{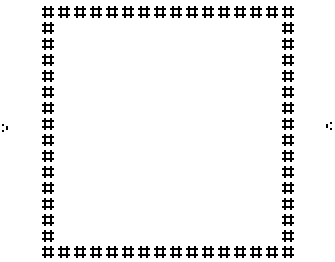}}} \hspace{0.55cm}
\subfigure[]{\scalebox{0.3}{\includegraphics{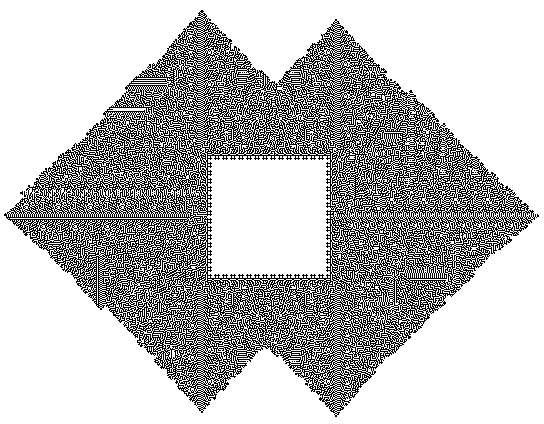}}}
\end{center}
\caption{Containment of growing pattern by indestructible localizations. (a) The explosion start from a reaction between two gliders, (b) final configuration of the container pattern, (c) Initial positions of gliders outside the box walled by indestructible patterns, (d) Interior of the box is protected from the growing pattern.}
\label{superNova}
\end{figure}

\begin{figure}
\begin{center}
\subfigure[]{\scalebox{0.35}{\includegraphics{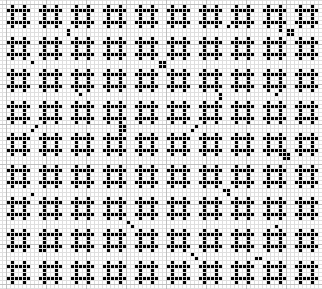}}} \hspace{0.2cm}
\subfigure[]{\scalebox{0.35}{\includegraphics{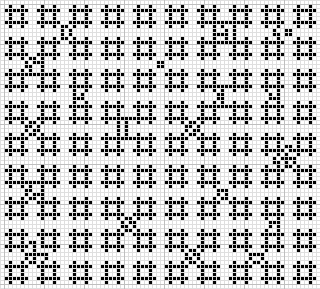}}}
\end{center}
\caption{Still life colony immune to local perturbations: (a)~initial configuration, where `viruses' are shown are irregular patterns of `1' states, (b)~final state demonstrates that initial local perturbations remain contained.}
\label{virus}
\end{figure}

The indestructible patterns discovered can be used to stop a 'supernova' explosions in Life-like rules. Usually a Life-like automaton development started at an arbitrary configuration exhibits unlimited growth (generally  related to nucleation phenomenon \cite{kn:Grav03}).

In rule $B2/S2345$ such an `uncontrollable' growth can be prevented by a regular arrangement of indestructible patterns. Examples are shown in Fig.~\ref{superNova}. In the first example (Fig.~\ref{superNova}a) two gliders collide inside a `box' made of still life patterns. The collision between the gliders lead to formation of growing pattern of `1' states. The propagation of the pattern is stopped by the indestructible localizations (Fig.~\ref{superNova}b). In the second example, gliders collide outside the box (Fig.~\ref{superNova}c)  however interior of the box remains resting (Fig.~\ref{superNova}c) due to impenetrable walls. Similarly, one can construct a colony of still life patterns immune to local perturbations. An example is shown in Fig.~\ref{virus}.

The indestructibility exemplified above allows us to use still life patterns to channel information in logical circuits. 

\section{Computing with propagating patterns}
\label{computing}

We built a computing scheme from channels --- long areas of `0'-state cells walled by still life blocks, and $T$-junctions\footnote{$T$-junction based control signals were suggested also in von Neumann~\cite{kn:von66} works.} --- sites where two or more channels join together. 

\begin{figure}
\centering
\includegraphics[width=0.35\textwidth]{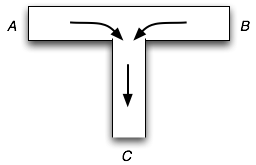}
\caption{$T$-shaped system processing information.}
\label{Tsystem}
\end{figure}

Each $T$-junction consists of two horizontal channels $A$ and $B$ (shoulders), acting as inputs, and a vertical channel, $C$, assigned as an output (Fig.~\ref{Tsystem}). Such type of circuitry has been already used to implement {\sc xor} gate in chemical laboratory precipitating reaction-diffusion systems~\cite{kn:ACA05}, and precipitating logical gates imitated in CA~\cite{kn:MAC08, kn:MAM08}. A minimal width of each channel equals three widths of the still life block (Fig.~\ref{basicPatterns}d) 
and width of a glider (Fig.~\ref{basicPatterns}a).

\begin{figure}[th]
\begin{center}
\subfigure[]{\scalebox{0.33}{\includegraphics{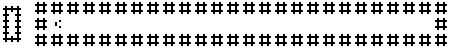}}} \hspace{0.7cm}
\subfigure[]{\scalebox{0.33}{\includegraphics{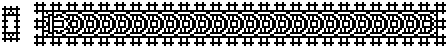}}} \hspace{0.7cm}
\subfigure[]{\scalebox{0.33}{\includegraphics{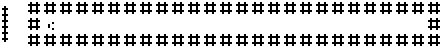}}} \hspace{0.7cm}
\subfigure[]{\scalebox{0.33}{\includegraphics{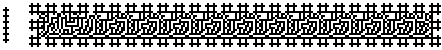}}}
\end{center}
\caption{Feedback channels constructed with still life patterns (fig. (a) and (c)) show the initial state with the empty channel and one glider respectively. The symmetric pattern represent value 0 (fig. (b)), and non-symmetric pattern represent value 1 (fig. (d)) late of glider reaction.}
\label{wave-comp-1}
\end{figure}

Boolean values are represented by reaction of gliders, positioned initially in the middle of channel, value 0 (Fig.~\ref{wave-comp-1} (a)), or slightly offset, value 1 (Fig.~\ref{wave-comp-1} (c)). The initial positions of the gliders determine outcomes of their reaction. Glider, corresponding to the value 0 is transformed to a regular symmetric pattern, similar to frozen waves of excitation activity (Fig.~\ref{wave-comp-1}b). Glider, representing  signal value 1, is transformed to transversally asymmetric patterns (Fig.~\ref{wave-comp-1}d). Both patterns propagate inside the channel with constant, advancing unit of channel length per step of discrete time.

\subsection{Implementation of logic gates and beyond}
\label{gates}

\begin{figure}[th]
\begin{center}
\subfigure[]{\scalebox{0.2}{\includegraphics{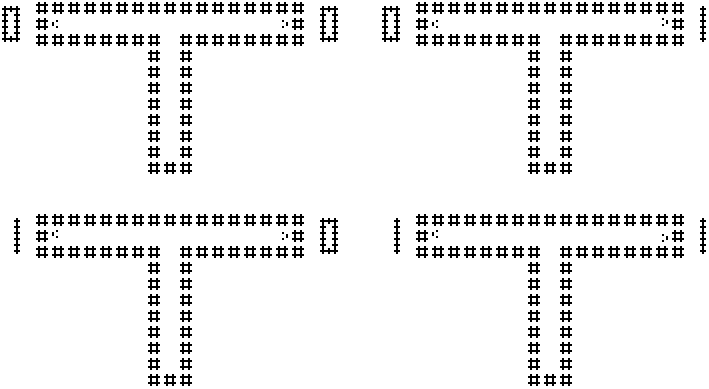}}} \hspace{0.7cm}
\subfigure[]{\scalebox{0.2}{\includegraphics{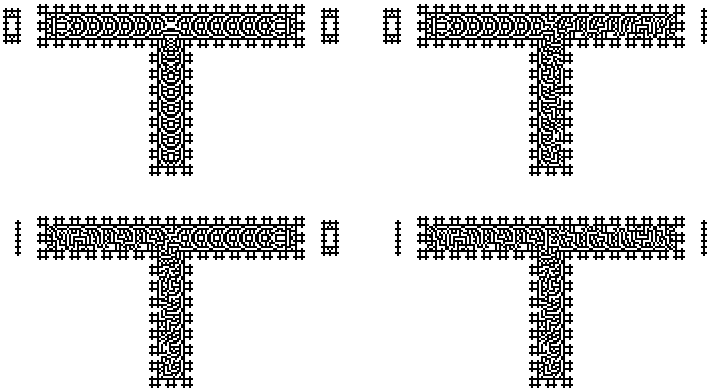}}} 
\subfigure[]{\scalebox{0.2}{\includegraphics{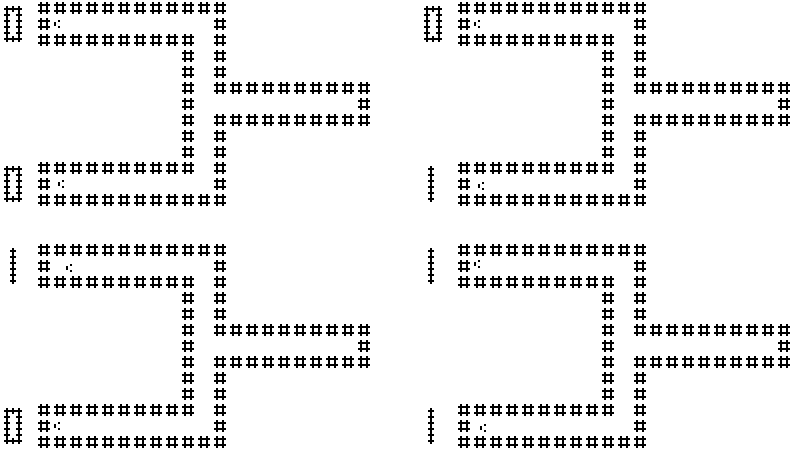}}} \hspace{0.2cm}
\subfigure[]{\scalebox{0.2}{\includegraphics{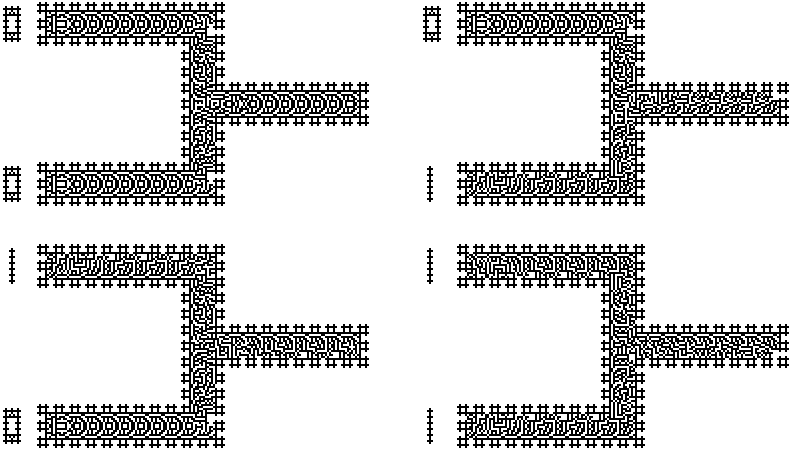}}}
\end{center}
\caption{Two kinds of {\sc or} gate implementation at the Life rule $B2/S2345$. Input binary values $a$ and $b$ they are represented as `In/0' or `In/1,' output result $c$ is represented by `Out/0' or `Out/1.' (a) and (c)~initial configurations of the gates, (d) and (e)~final configurations of the gates.}
\label{orGate}
\end{figure}

\begin{figure}
\begin{center}
\subfigure[]{\scalebox{0.2}{\includegraphics{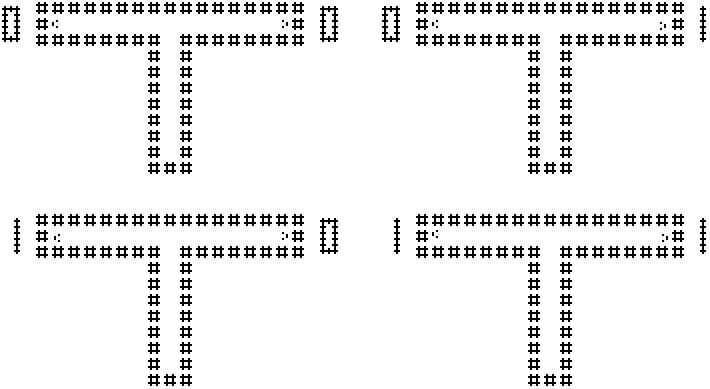}}} \hspace{0.7cm}
\subfigure[]{\scalebox{0.2}{\includegraphics{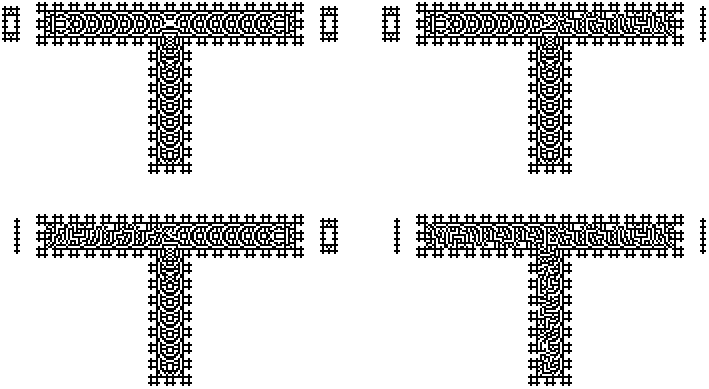}}} 
\subfigure[]{\scalebox{0.2}{\includegraphics{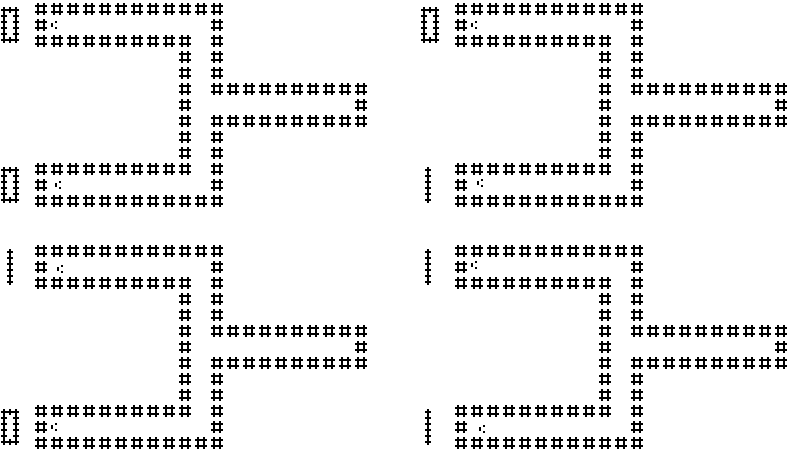}}} \hspace{0.2cm}
\subfigure[]{\scalebox{0.2}{\includegraphics{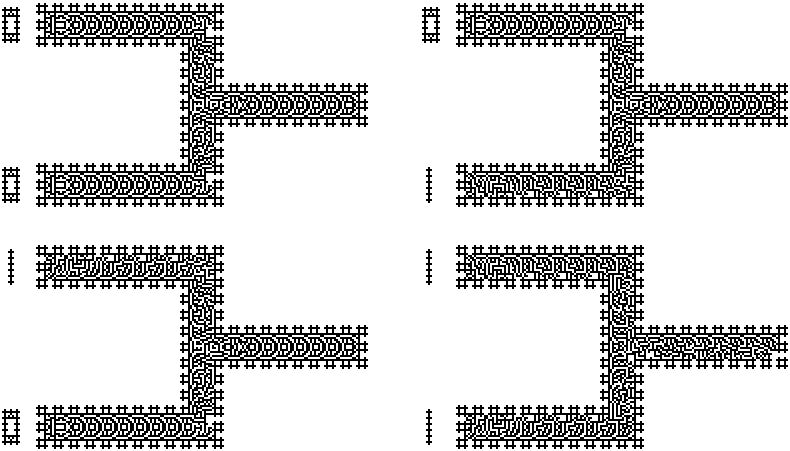}}}
\end{center}
\caption{Two kinds of {\sc and} gate implementation at the Life rule $B2/S2345$. (a) and (c)~initial configurations of the gates, (d) and (e)~final configurations of the gates.}
\label{andGate}
\end{figure}

\begin{figure}
\begin{center}
\subfigure[]{\scalebox{0.25}{\includegraphics{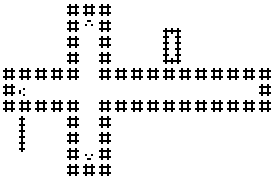}}} \hspace{0.8cm}
\subfigure[]{\scalebox{0.25}{\includegraphics{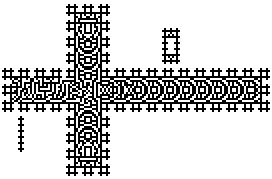}}} \\ 
\subfigure[]{\scalebox{0.25}{\includegraphics{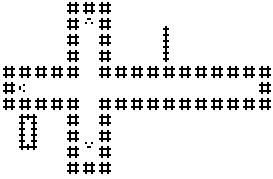}}} \hspace{0.8cm}
\subfigure[]{\scalebox{0.25}{\includegraphics{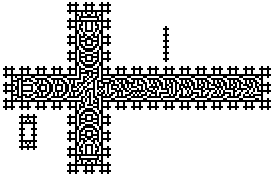}}}
\end{center}
\caption{{\sc not} gate implementation for input `1' (ab) and input `0' (cd). (a) and (c) are initial configurations, (b) and (d) are final configurations.}
\label{notGate-1}
\end{figure}

When patterns, representing values 0 and 1, meet at $T$-junctions they compete for the output channel. Depending on initial distance between gliders, one of the patterns wins and propagates along the output channel. Figures~\ref{orGate},~\ref{andGate} and~\ref{notGate-1} show final configurations of basic logical gates.

Figure~\ref{orGate} shows two implementations of {\sc or} gate. Due to different locations of gliders in initial configurations of gates, patterns in both implementations of gates are different however, results of computation are the same. Configurations of {\sc and} gate are shown in Fig.~\ref{andGate}.

\begin{figure}
\begin{center}
\subfigure[]{\scalebox{0.23}{\includegraphics{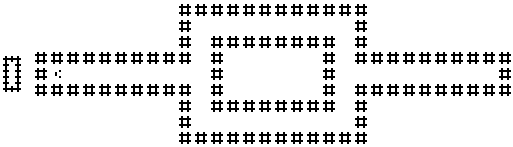}}} \hspace{0.5cm}
\subfigure[]{\scalebox{0.23}{\includegraphics{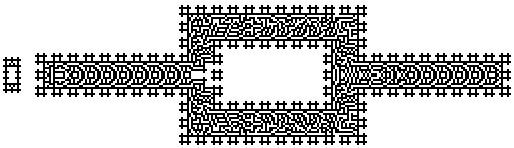}}} \hspace{0.5cm}
\subfigure[]{\scalebox{0.23}{\includegraphics{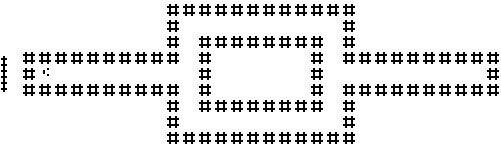}}} \hspace{0.5cm}
\subfigure[]{\scalebox{0.23}{\includegraphics{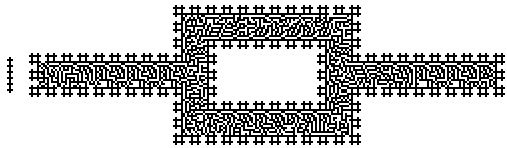}}}
\end{center}
\caption{Configurations of delay element for signal `0' (ab) and signal `1' (cd). (a) and (c) initial configurations, (b) and (d) final states.}
\label{delayDevice}
\end{figure}

Also we can implement a {\sc delay} element as shown in Fig.~\ref{delayDevice}.

\begin{figure}
\begin{center}
\subfigure[]{\scalebox{0.23}{\includegraphics{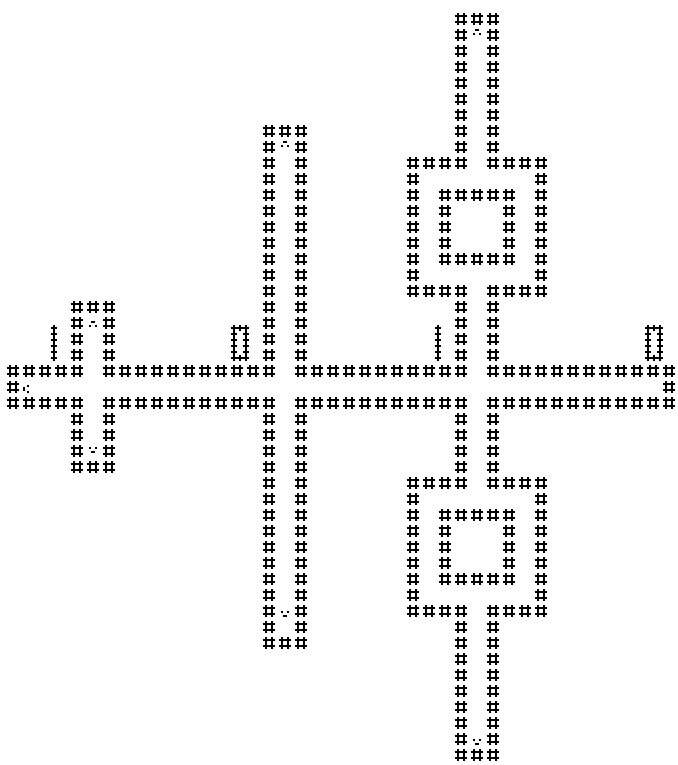}}} \hspace{0.1cm}
\subfigure[]{\scalebox{0.23}{\includegraphics{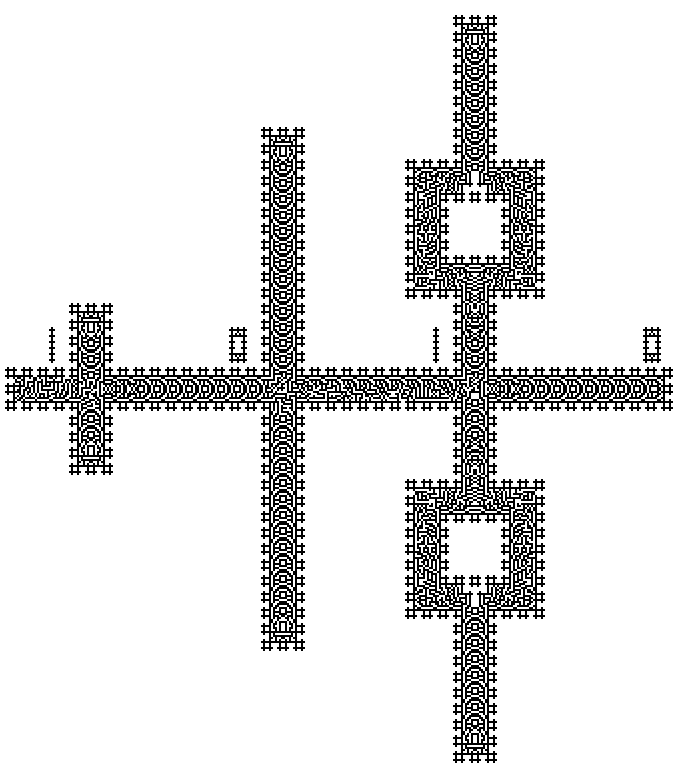}}}
\end{center}
\caption{Serial {\sc not} gate with delays.}
\label{notGate-2}
\end{figure}

The {\sc not} gate is implemented using additional channel, where control pattern is generated, propagate and interfere with data-signal pattern. Initial and final configurations of {\sc not} gate are shown in Fig.~\ref{notGate-1}. Using delay elements we can construct serial channels with any number of {\sc not} gates (Fig.~\ref{notGate-2}). Number of control channels growth proportionally to number of gates in the circuit. Of course, it could not be the most elegant and efficient way of constructing {\sc not} gate, but useful for our purposes in $B2/S2345$.

\subsection{Majority gate}
\label{majGates}

\begin{figure}[th]
\begin{center}
\subfigure[]{\scalebox{0.37}{\includegraphics{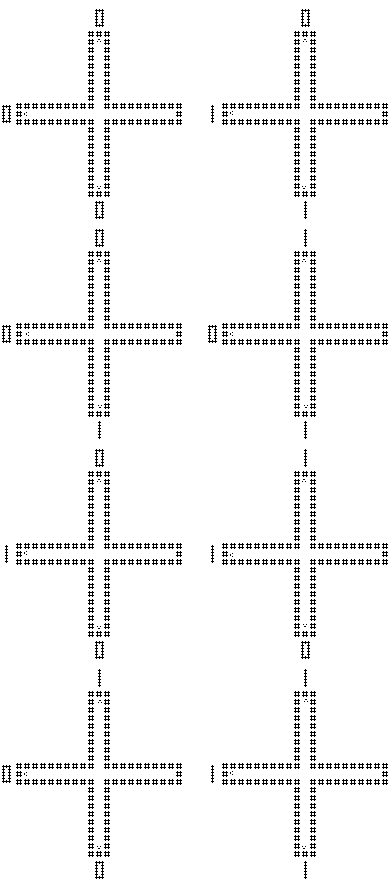}}} \hspace{0.7cm}
\subfigure[]{\scalebox{0.37}{\includegraphics{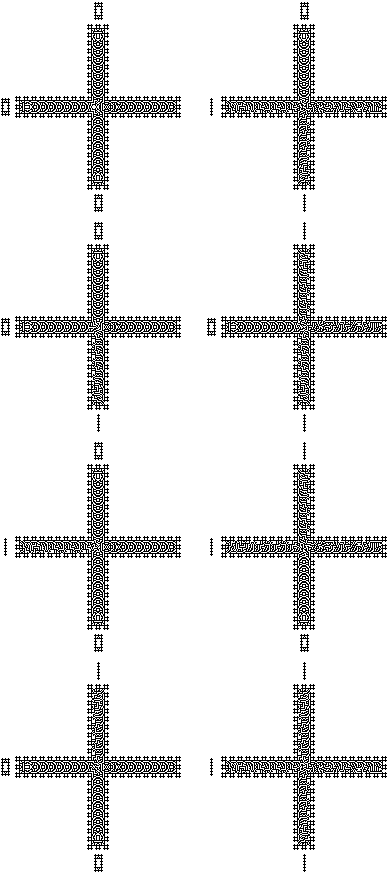}}}
\end{center}
\caption{{\sc Majority} gate implementation $(a \wedge b) \vee (a \wedge c) \vee (b \wedge c)$: (a)~initial configuration: majority input values In/0 (first column), and majority input values In/1 (second column), and (b)~final configurations of the majority gates.}
\label{majorityGate}
\end{figure}

Implementation of {\sc majority} gate is shown in Fig.~\ref{majorityGate}. The gate has three inputs: North, West and South channels, and one output: East channel. Three propagating pattern, which represent inputs, collide at the cross-junction of the gate. The resultant pattern is recorded at the output channel. Similarly gates in quantum-dot cellular automata  are designed~\cite{kn:Por99}.

\subsection{Implementation of binary adders}
\label{adders}

We represent two ways of implementing partial and full binary adders. First we can consider design based on cascading of logic gates, then  second design employing only {\sc not-majority} gates.

\subsubsection{Adder via cascading of logical gates}
\label{adderCascade}

\begin{figure}[th]
\begin{center}
\subfigure[]{\scalebox{0.46}{\includegraphics{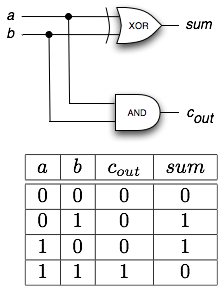}}}  \hspace{0.3cm}
\subfigure[]{\scalebox{0.46}{\includegraphics{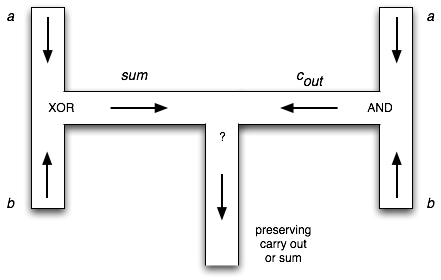}}}\end{center}
\caption{(a)~Half-adder circuit and its respective true table, and (b)~scheme of half-adder implementation in geometrically constrained medium.}
\label{adderHalf-1}
\end{figure}

Conventional logic circuit and truth table of a binary half-adder are shown in Fig.~\ref{adderHalf-1}a (true tables were derived from \cite{kn:MAM08}).\footnote{\url{http://uncomp.uwe.ac.uk/genaro/Diffusion_Rule/B2-S2345678.html}} Schematic representation of the half-adder via $T$-junctions between channels is shown in Fig.~\ref{adderHalf-1}b. 

\begin{figure}
\begin{center}
\subfigure[]{\scalebox{0.27}{\includegraphics{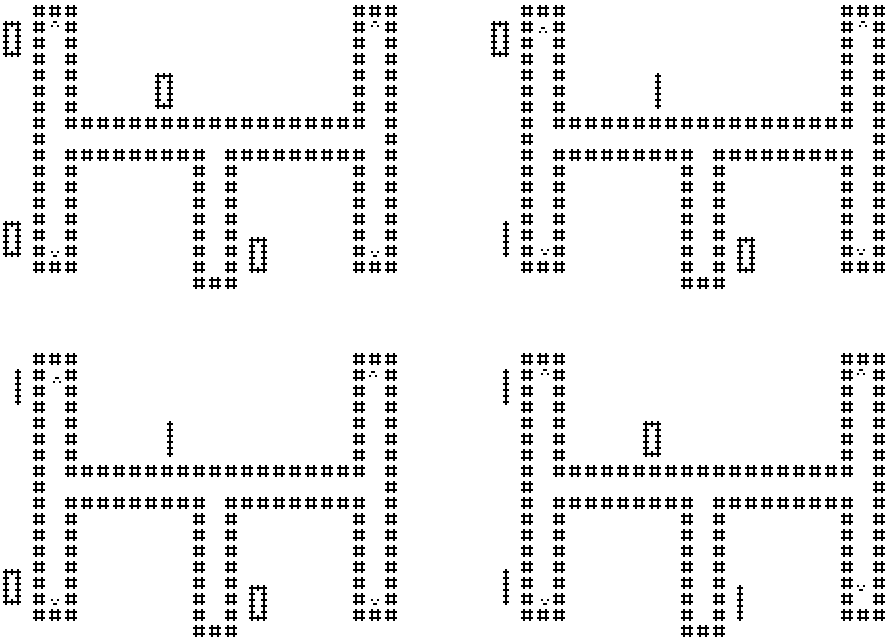}}}  
\subfigure[]{\scalebox{0.27}{\includegraphics{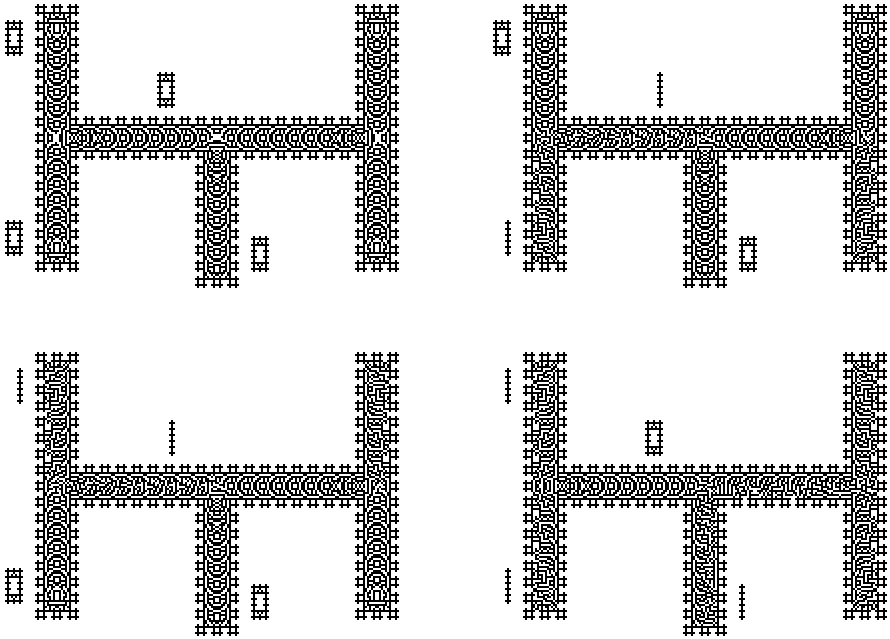}}}\end{center}
\caption{Half adder implemented in rule $B2/S2345$. Configurations represent sums $0+0$, $0+1$, $1+0$, and $1+1$ respectively, carry-out is preserved.}
\label{adderHalf-2}
\end{figure}

Final configurations of patterns in a one-bit half-adder are calculated for all inputs values. Figure~\ref{adderHalf-2}a shows the initial configuration of inputs and outputs. Resultant patterns are shown in Fig.~\ref{adderHalf-2}b where the last patterns is a carry-out operation for the  next half-adder.
 
\begin{figure}[th]
\centering
\includegraphics[width=1\textwidth]{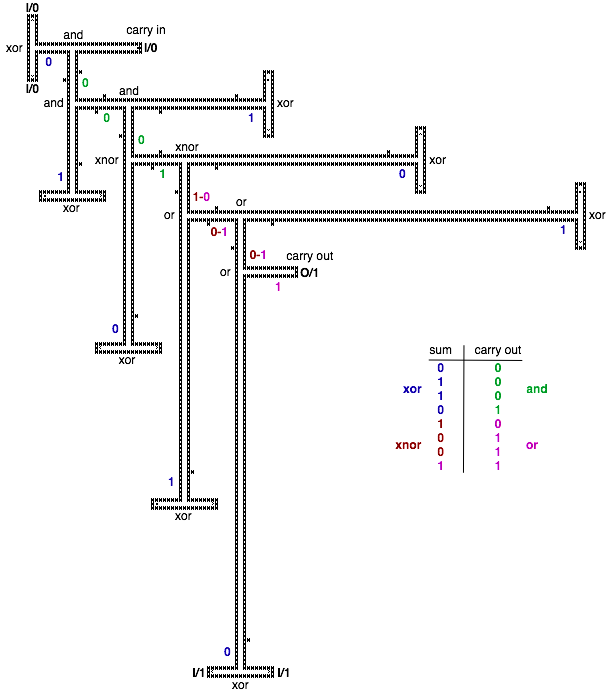}
\caption{Configuration and description of full binary adder.}
\label{fullAdder-1}
\end{figure}

The circuit can be extended to a full binary adder via cascading of logical gates \cite{kn:GQZ07}. Configuration of the adder, built of still life blocks, and its description are shown in Fig.~\ref{fullAdder-1}. The full adder consists of 16 $T$-junctions, linked together by channels and involve synchronization signals.

\begin{figure}[th]
\centering
\includegraphics[width=1\textwidth]{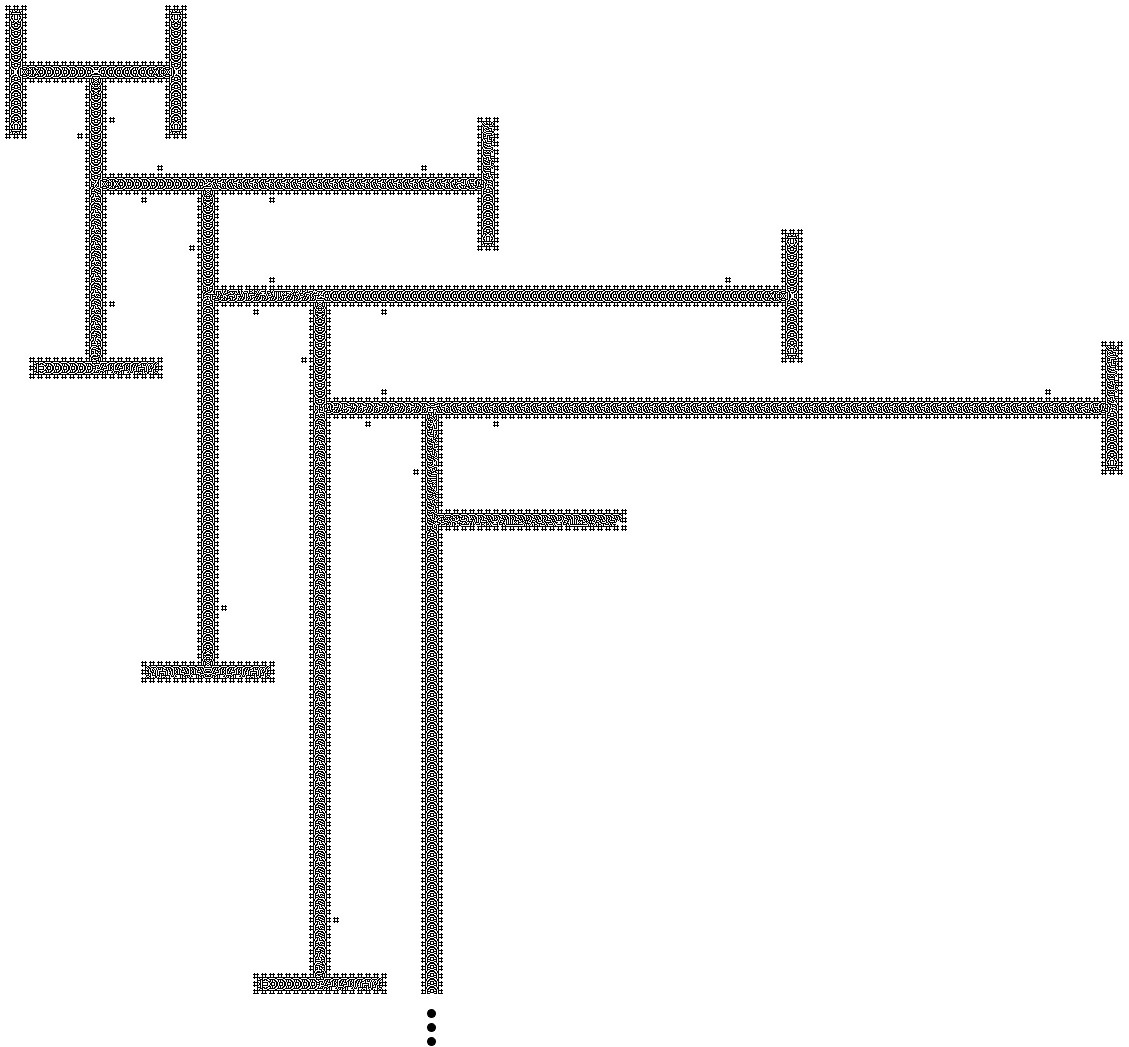}
\caption{Zoomed-in data-area of the serial implementation of full binary adder.}
\label{fullAdder-2}
\end{figure}

Working prototype of full adder is constructed in $1,118 \times 1,326$ cells lattice. In total, initial configuration has a population of 28,172 alive cells.  The prototypes working cycle is 1,033 time steps with a final population of 63,662 alive cells. A data-area of the full adder is shown in Fig.~\ref{fullAdder-2}.

\subsubsection{Construction of adder via {\sc not-majority} gates}

The second model represents a binary adder constructed of three {\sc not-majority} gates and two inverters. Such type of adder appears in several publications, particularly in construction of the arithmetical circuits in quantum-dot cellular automata~\cite{kn:Por99}. Original version of the adder using {\sc not-majority} gates was suggested by Minsky in his designs of artificial neural networks~\cite{kn:Mins67}. 

\begin{figure}[th]
\centering
\includegraphics[width=0.53\textwidth]{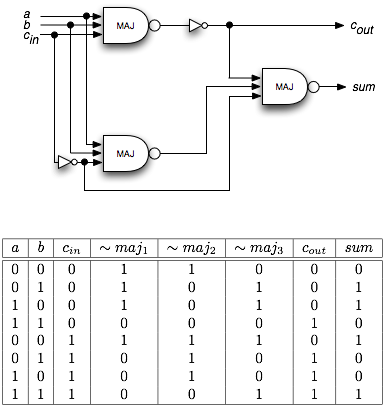}
\caption{Circuit and truth table of a full binary adder comprised of {\sc not-majority} gates and inverters.}
\label{fullAdder-3}
\end{figure}

\begin{figure}[th]
\centering
\includegraphics[width=0.54\textwidth]{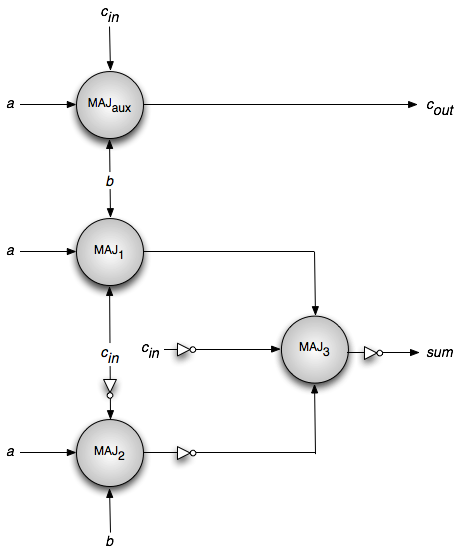}
\caption{Schematic diagram of a full binary adder comprised of {\sc not-majority} gates. Delay elements are not shown.}
\label{fullAdder-3a}
\end{figure}

Figure~\ref{fullAdder-3} shows the classic circuit and the true table illustrating the dynamics of this adder. This way, figure~\ref{fullAdder-3a} represents a scheme of the adder to implement in $B2/S2345$. The scheme highlights critical points where some extra gates are necessary to adjust inputs and synchronize times of collisions.

\begin{figure}
\centering
\includegraphics[width=0.7\textwidth]{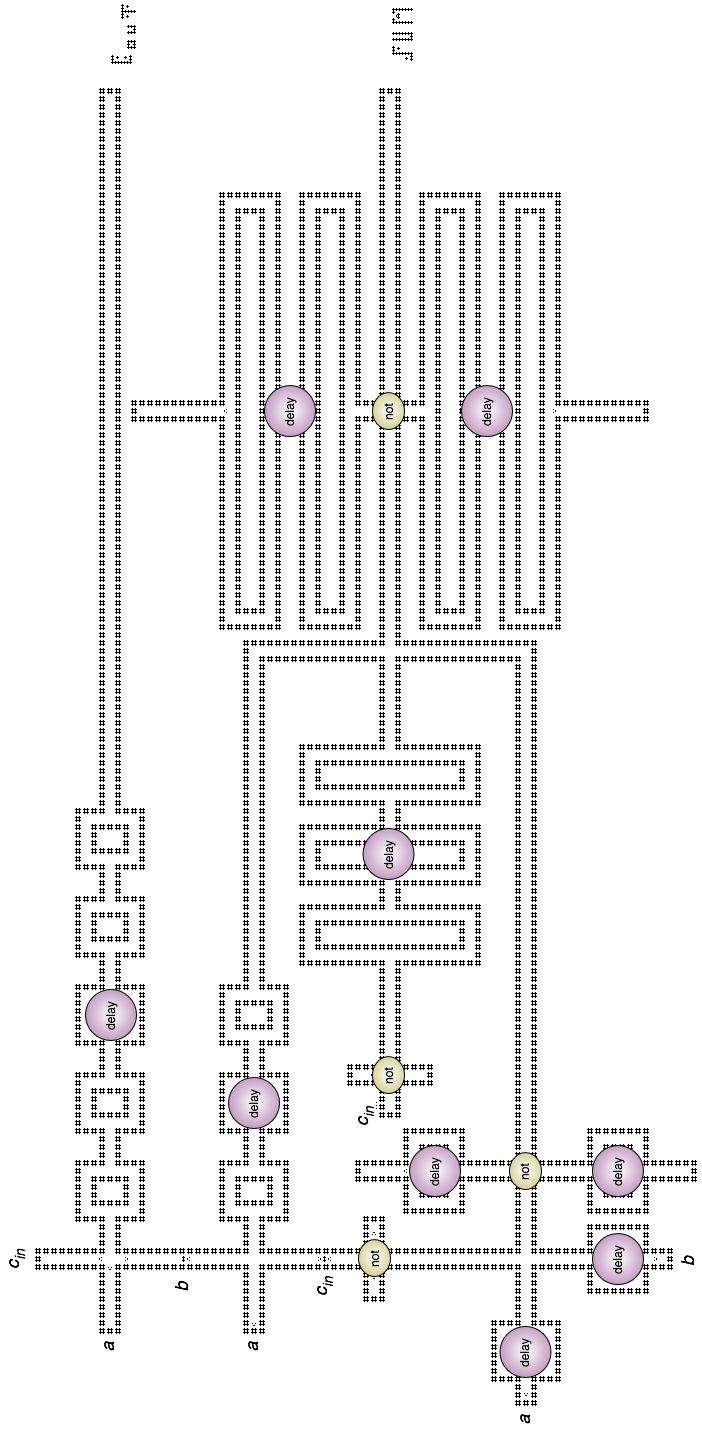}
\caption{Full binary adder designed with {\sc not-majority} gates, stages and main circuit implementation on $B2/S2345$ evolution space.}
\label{fullAdder-4}
\end{figure}

\begin{figure}
\centering
\includegraphics[width=0.7\textwidth]{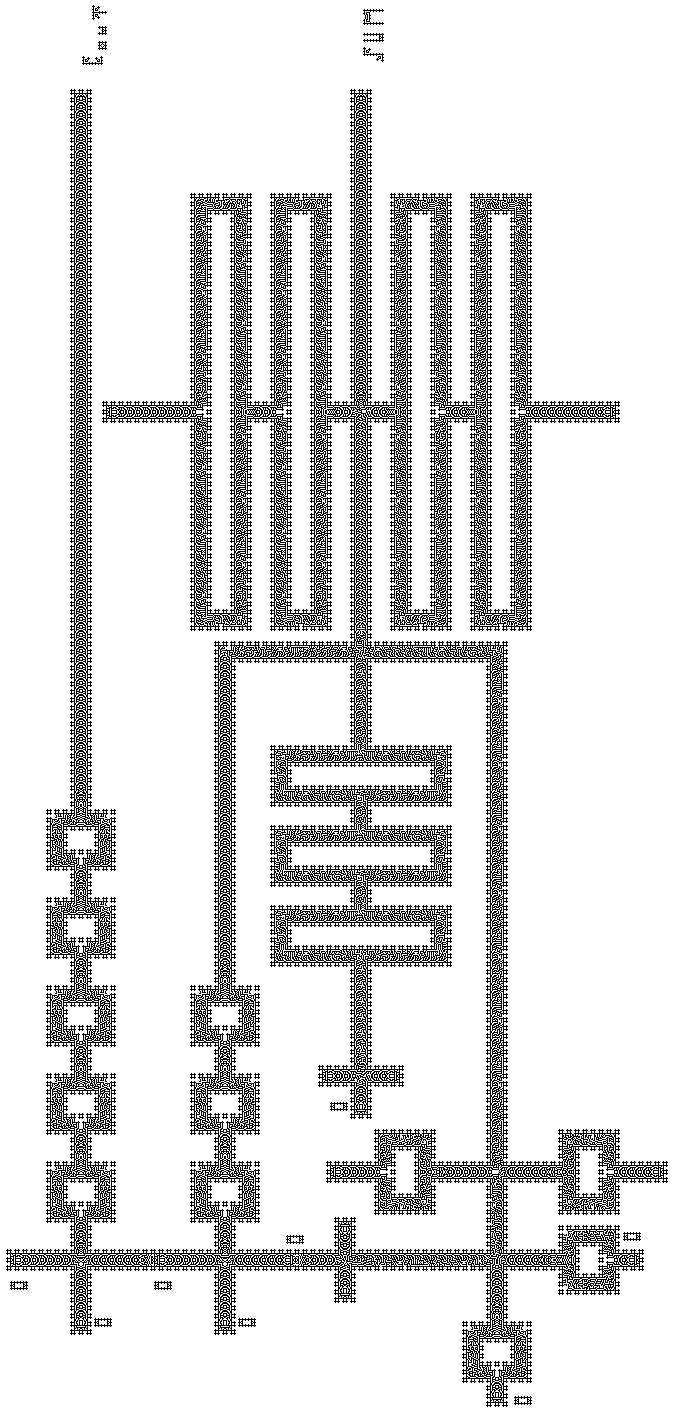}
\caption{Final configuration of the adder for inputs $a=0$, $b=0$ and $c_{in}=0$, and outputs $c_{out}=0$ and $sum=0$.}
\label{fullAdder-5}
\end{figure}

\begin{figure}
\centering
\includegraphics[width=0.7\textwidth]{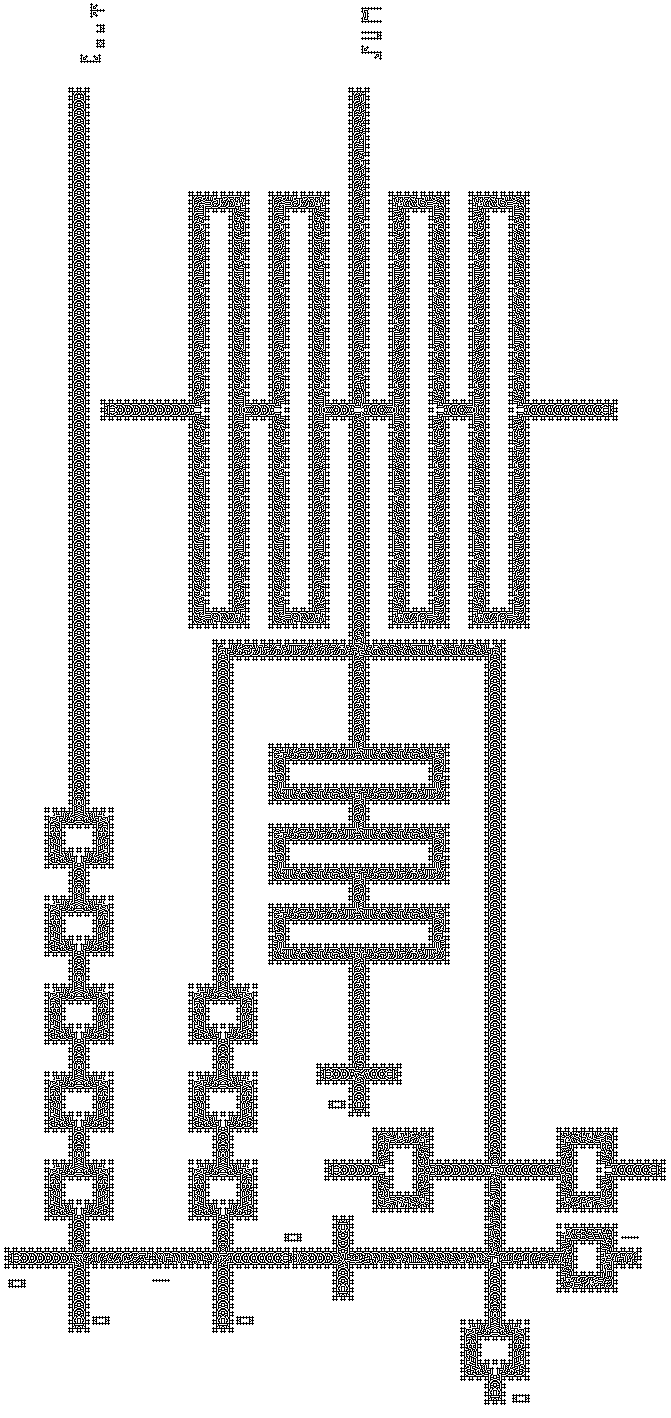}
\caption{Configuration of the adder for inputs $a=0$, $b=1$ and $c_{in}=0$, and outputs $c_{out}=0$ and $sum=1$.}
\label{fullAdder-6}
\end{figure}

\begin{figure}
\centering
\includegraphics[width=0.7\textwidth]{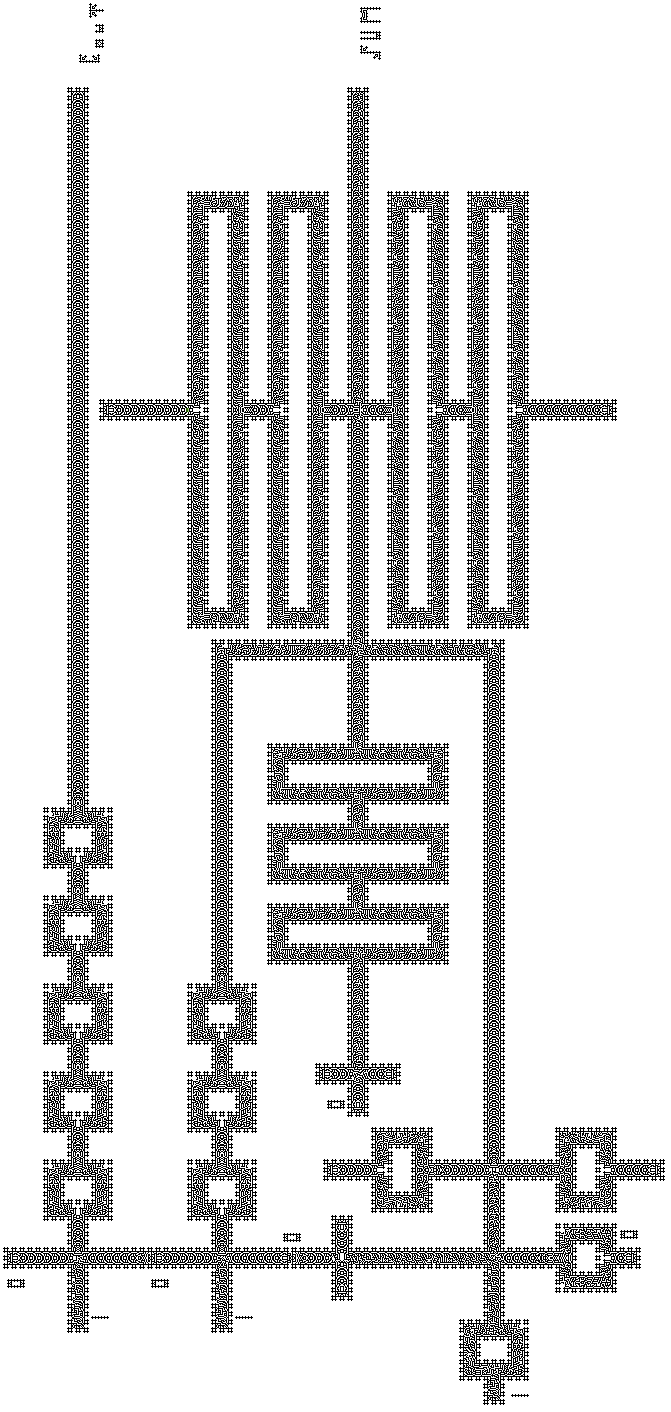}
\caption{Configuration of the adder for inputs  $a=1$, $b=0$ and $c_{in}=0$, and outputs $c_{out}=0$ and $sum=1$.}
\label{fullAdder-7}
\end{figure}

\begin{figure}
\centering
\includegraphics[width=0.7\textwidth]{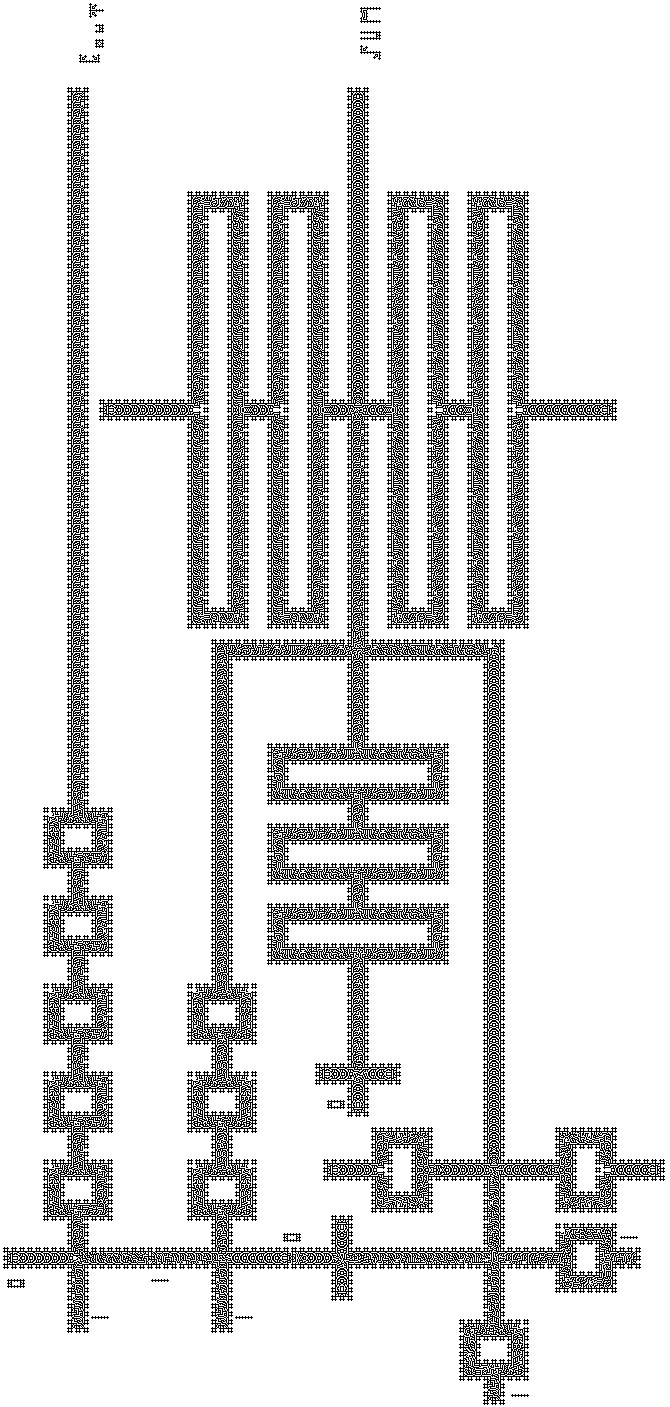}
\caption{Configuration of the adder for inputs  $a=1$, $b=1$ and $c_{in}=0$, and outputs $c_{out}=1$ and $sum=0$.}
\label{fullAdder-8}
\end{figure}

\begin{figure}
\centering
\includegraphics[width=0.7\textwidth]{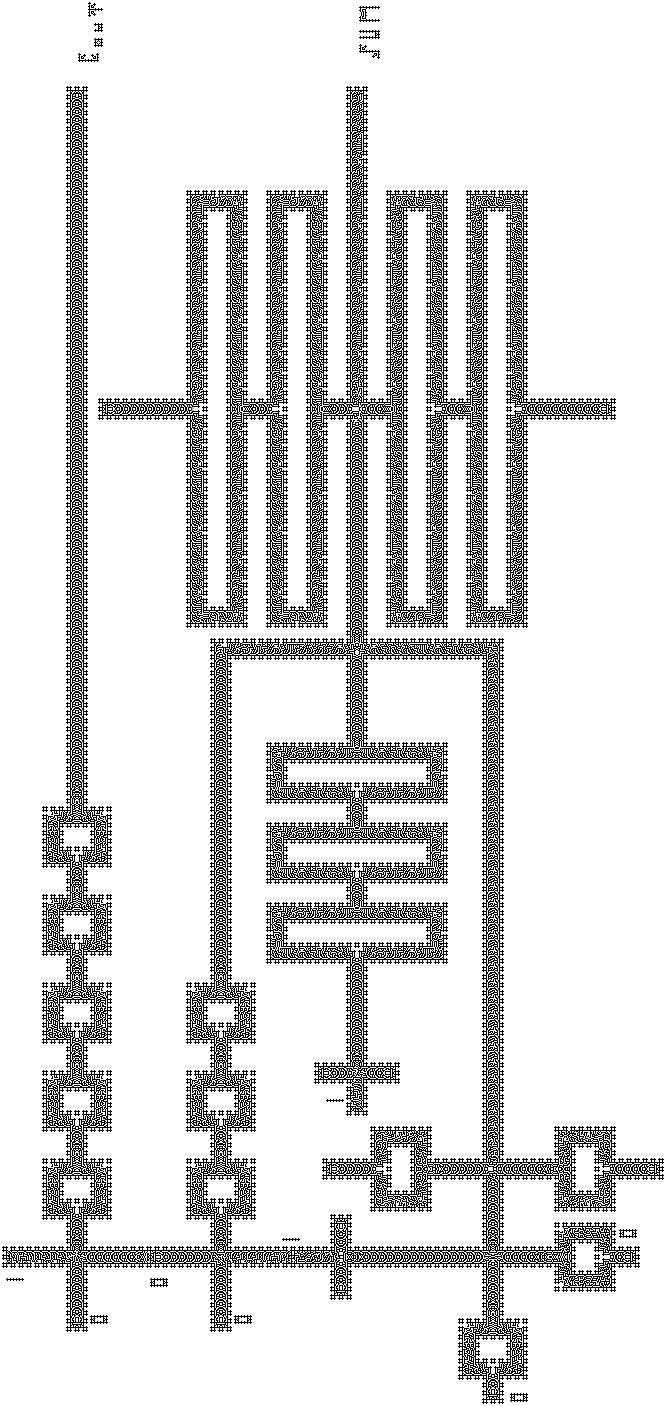}
\caption{Configuration of the adder for inputs $a=0$, $b=0$ and $c_{in}=1$, and outputs $c_{out}=0$ and $sum=1$.}
\label{fullAdder-9}
\end{figure}

\begin{figure}
\centering
\includegraphics[width=0.7\textwidth]{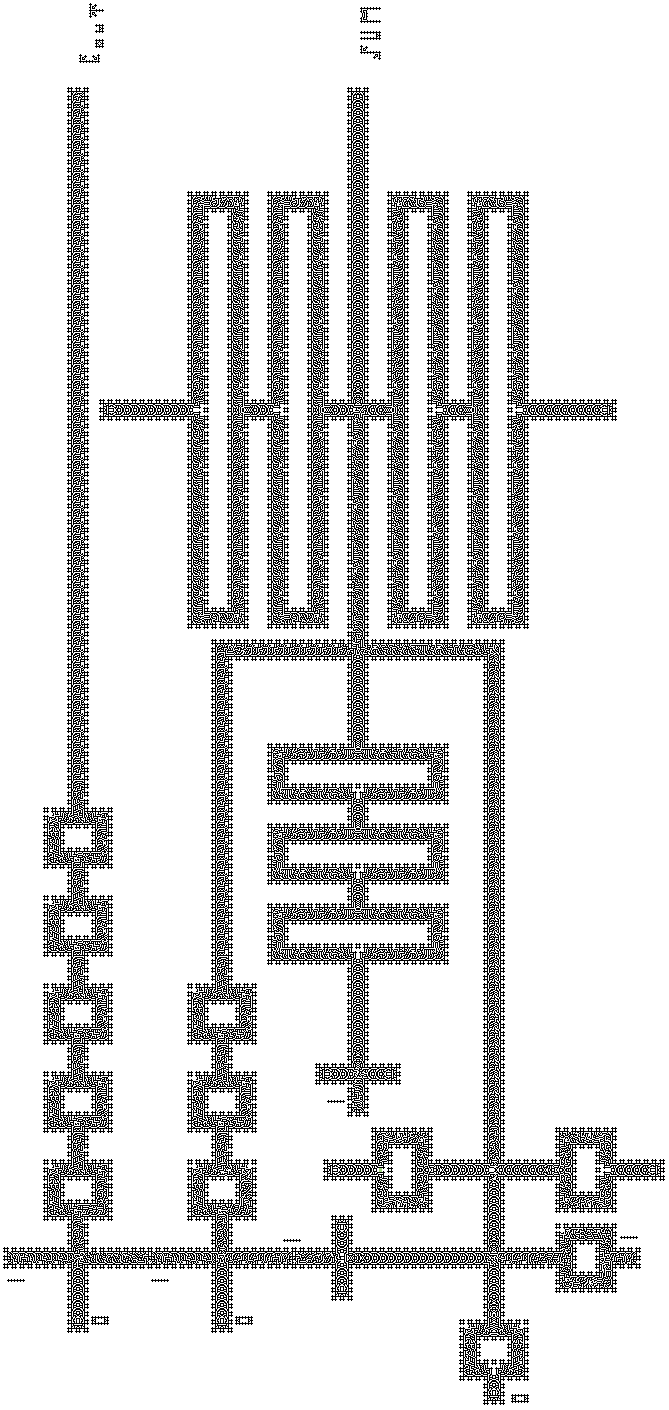}
\caption{Configuration of the adder for inputs $a=0$, $b=1$ and $c_{in}=1$, and outputs $c_{out}=1$ and $sum=0$.}
\label{fullAdder-10}
\end{figure}

\begin{figure}
\centering
\includegraphics[width=0.7\textwidth]{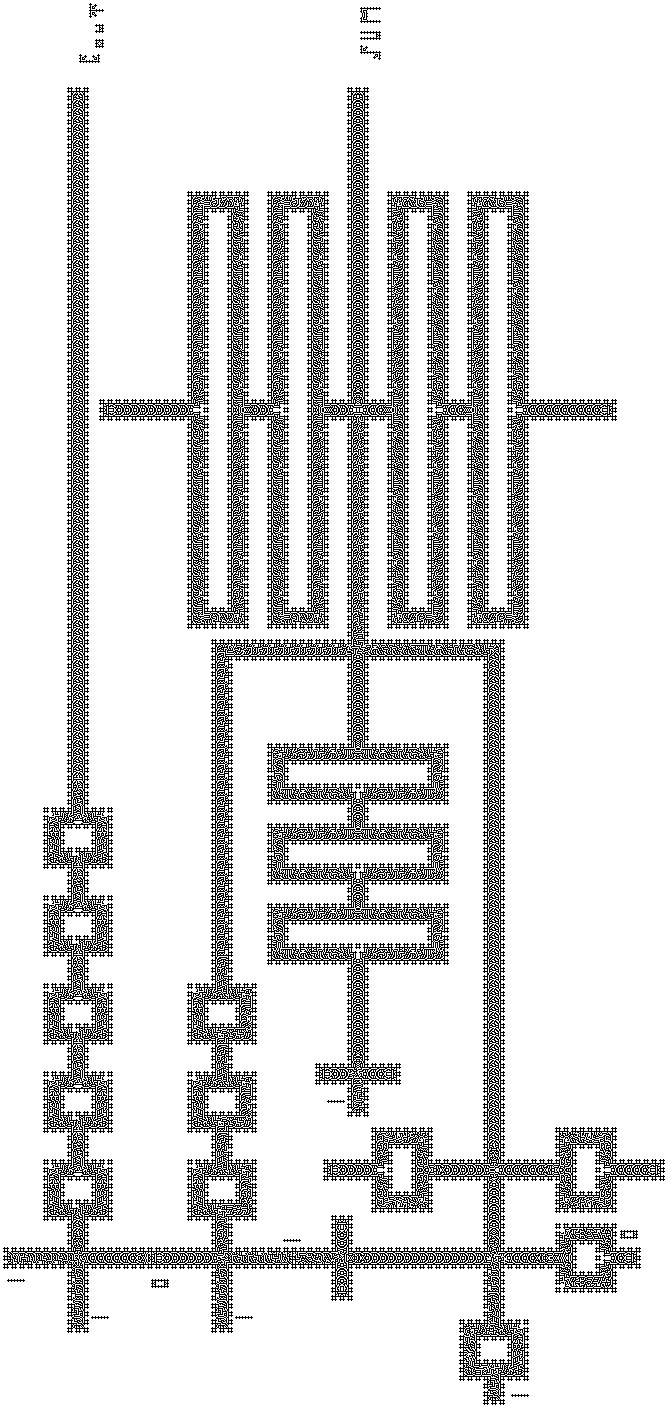}
\caption{Configuration of the adder for inputs $a=1$, $b=0$ and $c_{in}=1$, and outputs $c_{out}=1$ and $sum=0$.}
\label{fullAdder-11}
\end{figure}

\begin{figure}
\centering
\includegraphics[width=0.7\textwidth]{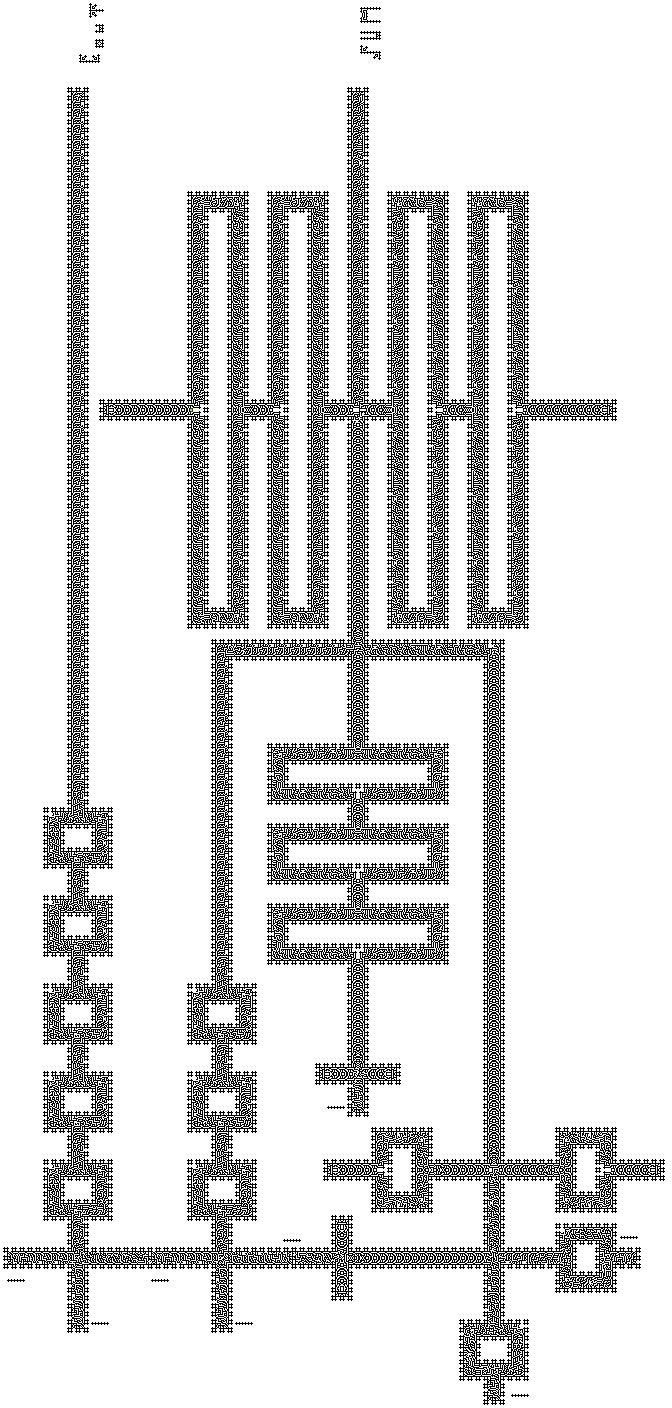}
\caption{Configuration of the adder for inputs $a=1$, $b=1$ and $c_{in}=1$, and outputs $c_{out}=1$ and $sum=1$.}
\label{fullAdder-12}
\end{figure}

Figure~\ref{fullAdder-4} presents most important stages of the full adder on $B2/S2345$ evolution space standing out {\sc delays} stages and {\sc not} gates. The adder is implemented in $1,402 \times 662$ lattice that relates an square of 928,124 cells lattice with an initial population of 56,759 cells in state `1.' Final configurations of the adder for every initial configuration of inputs are shown in Figs.~\ref{fullAdder-5}--\ref{fullAdder-12} with a final population of 1,439 alive cells on an average of 129,923 generations.

\section{Conclusions}
\label{conclusions}

We studied a cellular automaton model --- Life-like rule $B2/S2345$ -- of a precipitating chemical system. We demonstrated that chaotic rule $B2/S2345$ supports stationary (still lifes) and mobile (gliders and propagating patterns) localizations. That relates another case where a chaotic rule contains non evident complex behavior and how such systems could have some computing information on its evolution \cite{kn:MAM, kn:Mit01}.

We have shown how construct basic logical gates and arithmetical circuits by restricting propagation of patters in a channels made of indestructible stationary localizations. Disadvantage of the approach presented is that computing space is geometrically constrained and not all cells of automaton lattices are used in computation. However, the geometrical constraining brings some benefits as well. Most computing circuits in Life-like automata are built using very complex dynamics of collisions between gliders and still lifes~\cite{kn:Chap02, kn:Ren02, kn:Gou09}, in our case gliders are used only to `ignite' propagating patterns in the channels~\cite{kn:ACA05, kn:Wain73}. In future studies we are planning to implement the computing architecture designed in the paper to manufacture experimental prototypes of precipitating chemical computers; they will be based on crystallization of `hot ice'~\cite{kn:Ada09}.

Implementations and constructions are done in {\it Golly system} (\url{http://golly.sourceforge.net/}). Source configurations and specific initial condition (RLE files) to reproduce the results are available in  \url{http://uncomp.uwe.ac.uk/genaro/Life_dc22.html}.

\section*{Acknowledgement}
Genaro J. Mart{\'i}nez was partially funded by Engineering and Physical Sciences Research Council (EPSRC), United Kingdom, grant EP/F054343, and postdoctoral funding at the ICN and C3 of UNAM. Kenichi Morita was partially funded by Grant-in-Aid for Scientific Research (C) No. 21500015 from JSPS.



\begin{thebibliography}{99}

\bibitem{kn:Ada01} A. Adamatzky: {\em Computing in Nonlinear Media and Automata Collectives}, Institute of Physics Publishing, Bristol and Philadelphia (2001).

\bibitem{kn:Ada02} A. Adamatzky (ed.): {\em Collision-Based Computing}, Springer (2002).

\bibitem{kn:ACA05} A. Adamatzky, B. L. Costello, T. Asai: {\em Reaction-Diffusion Computers}, Elsevier (2005).

\bibitem{kn:AMS05} A. Adamatzky, G. J. Mart\'{\i}nez, J. C. Seck-Tuoh-Mora: Phenomenology of reaction-diffusion binary-state cellular automata, {\em Int. J. Bifurcation and Chaos} {\bf 16 (10)} 1--21 (2006).

\bibitem{kn:Ada07} A. Adamatzky: Physarum machines: encapsulating reaction-diffusion to compute spanning tree, {\em Naturwisseschaften} {\bf 94} 975--980 (2007).

\bibitem{kn:Ada09} A. Adamatzky: Hot ice computer, {\em Physics Letters A} {\bf 374(2)} 264--271 (2009).

\bibitem{kn:BCG82} E. R. Berlekamp, J. H. Conway, R. K. Guy: {\em Winning Ways for your Mathematical Plays}, Academic Press, (vol. 2, chapter 25) (1982).

\bibitem{kn:BE03} V. Beato, H. Engel: Pulse propagation in a model for the photosensitive Belousov-Zhabotinsky reaction with external noise. In: {\em Noise in Complex Systems and Stochastic Dynamics}, Edited by Schimansky-Geier~L., Abbott~D., Neiman~A., Van~den~Broeck~C. Proc. SPIE (5114) 353--362 (2003).

\bibitem{kn:Chap02} P. Chapman: Life Universal Computer, \url{http://www.igblan.free-online.co.uk/igblan/ca/} (2002).

\bibitem{kn:CM91} H. Chat\'e, P. Manneville: Evidence of collective behavior in cellular automata, {\em Europhysics Letters} {\bf 14} 409--413 (1991).

\bibitem{kn:Cook03} M. Cook: Still Life Theory, In \cite{kn:GM03}, 93--118 (2003).

\bibitem{kn:CTS09} B. L. Costello, R. Toth, C. Stone, A. Adamatzky, L. Bull: Implementation of Glider guns in the light sensitive Belousov-Zhabotinsky medium (2009), {\em Physical Review E} {\bf 79}  026114 (2009).

\bibitem{kn:DAK98} C. Dupont, K. Agladze, V. Krinsky: Excitable medium with left-right symmetry breaking, {\em Physica A} {\bf 249} 47--52 (1998).

\bibitem{kn:Gard70} M. Gardner: Mathematical Games --- The fantastic combinations of John H. Conway's new solitaire game Life, {\em Scientific American} {\bf 223} 120--123 (1970).

\bibitem{kn:GG03} J. Gorecka, J. Gorecki: T-shaped coincidence detector as a band filter of chemical signal frequency, {\em Phys. Rev. E} {\bf 67} 067203 (2003).

\bibitem{kn:GM96} D. Griffeath, C. Moore:  Life Without Death is P-complete, {\it Complex Systems} {\bf 10} 437--447 (1996).

\bibitem{kn:GM03} D. Griffeath, C. Moore (eds.): {\em New constructions in cellular automata}, Oxford University Press (2003).

\bibitem{kn:Gou09} A. Goucher: Completed Universal Computer/Constructor (2009). In: \url{http://pentadecathlon.com/lifeNews/2009/08/post.html}.

\bibitem{kn:GQZ07} Z. Guan, X. Qin, Y. Zhang, Q. Shi: Network Structure Cascade for Reversible Logic, {\em Proceedings of the Third International Conference on Natural Computation} {\bf 3} 306--310 (2007).

\bibitem{kn:Grav03} J. Gravner: Growth Phenomena in Cellular Automata, In \cite{kn:GM03}, 161--181 (2003).

\bibitem{kn:GV87} H. A. Gutowitz, J. D. Victor: Local structure theory in more that one dimension, {\em Complex Systems} {\bf 1} 57--68 (1987).

\bibitem{kn:GYI03} J. Gorecki, K. Yoshikawa, Y. Igarashi: On chemical reactors which can count, {\em J. Phys. Chem. A} {\bf 107} 1664--1669 (2003).

\bibitem{kn:IM00} K. Imai, K. Morita: A computation-universal two-dimensional 8-state triangular reversible cellular automaton, {\em Theoret. Comput. Sci.} {\bf 231} 181--191 (2000).

\bibitem{kn:KPK90} H. J. Krug, L. Pohlmann, L. Kuhnert: Analysis of the modified complete Oregonator (MCO) accounting for oxygen- and photosensitivity of Belousov-Zhabotinsky systems, {\em J. Phys. Chem.} {\bf 94} 4862--4866 (1990).

\bibitem{kn:KYA97} T. Kusumi, T. Yamaguchi, R. Aliev, T. Amemiya, T. Ohmori, H. Hashimoto, K. Yoshikawa: Numerical study on time delay for chemical wave transmission via an inactive gap, {\em Chem. Phys. Lett.} {\bf 271} 355--60 (1997).

\bibitem{kn:MAC08} G. J. Mart\'{\i}nez, A. Adamatzky, B. L. Costello: On logical gates in precipitating medium: cellular automaton model, {\em Physics Letters A} {\bf 1 (48)} 1--5 (2008).

\bibitem{kn:MAM} G. J. Mart\'{\i}nez, A. Adamatzky, H. V. McIntosh: Localization dynamic in a binary two-dimensional cellular automaton: the Diffusion Rule, {\em Journal of Cellular Automata} {\bf 5(4-5)} 289--313 (2010).

\bibitem{kn:MAM08} G. J. Mart\'{\i}nez, A. Adamatzky, H. V. McIntosh, B. L. Costello: Computation by competing patterns: Life rule $B2/S2345678$, In {\em Automata 2008: Theory and Applications of Cellular Automata}, Adamatzky, A. {\em et. al} (eds.), Luniver Press (2008).

\bibitem{kn:Mc88} H. V. McIntosh: Life's Still Lifes, \url{http://delta.cs.cinvestav.mx/~mcintosh} (1988).

\bibitem{kn:Mc90} H. V. McIntosh: Wolfram's Class IV and a Good Life, {\em Physica D} {\bf 45} 105--121 (1990).

\bibitem{kn:Mins67} M. Minsky: {\em Computation: Finite and Infinite Machines}, Prentice Hall (1967).

\bibitem{kn:Mit01} M. Mitchell: Life and evolution in computers,  {\em History and Philosophy of the Life Sciences} {\bf 23} 361--383 (2001).

\bibitem{kn:MLH97} M. Magnier, C. Lattaud, J.-K. Heudin: Complexity Classes in the Two-dimensional Life Cellular Automata Subspace, {\em Complex Systems} {\bf 11 (6)} 419--436 (1997).

\bibitem{kn:MMI99} K. Morita, M. Margenstern, K. Imai: Universality of reversible hexagonal cellular automata, {\em Theoret. Informatics Appl.} {\bf 33} 535--550 (1999).

\bibitem{kn:MMZ05} G. J. Mart\'{\i}nez, A. M. M\'endez, M. M. Zambrano:  Un subconjunto de aut\'omata celular con comportamiento complejo en dos dimensiones, \url{http://uncomp.uwe.ac.uk/genaro/Papers/Papers_on_CA.html} (2005).

\bibitem{kn:MYI01} I. N. Motoike, K. Yoshikawa, Y. Iguchi, S. Nakata: Real-time memory on an excitable field, {\em Phys. Rev. E} {\bf 63} 036220 (2001).

\bibitem{kn:Por99} W. Porod, C. S. Lent, G. H. Bernstein, A. O. Orlov, I. Amlani, G. L. Snider, J. L. Merz: Quantum-dot cellular automata: computing with coupled quantum dots, {\it Int. J. Electronics} {\bf 86 (5)} 549--590 (1999).

\bibitem{kn:PW85} N. Packard, S. Wolfram: Two-dimensional cellular automata, {\em J. Statistical Physics} {\bf 38} 901--946 (1985).

\bibitem{kn:Ren02} P. Rendell: Turing universality of the game of life, In \cite{kn:Ada02}, 513--540 (2002).

\bibitem{kn:Ren03} J. P. Rennard: Implementation of Logical Functions in the Game of Life, In \cite{kn:Ada02}, 491--512 (2002).

\bibitem{kn:SG01} J. Sielewiesiuk, J. Gorecki: Logical functions of a cross junction of excitable chemical media, {\em J. Phys. Chem. A} {\bf 105} 8189--95 (2001).

\bibitem{kn:Toff98} T. Toffoli: Non-Conventional Computers, {\em Encyclopedia of Electrical and Electronics Engineering} (John Webster Ed.) {\bf 14} 455--471, Wiley \& Sons, (1998).

\bibitem{kn:TS95} A. T\'{o}th, K. Showalter: Logic gates in excitable media, {\em J. Chem. Phys.} {\bf 103} 2058-66 (1995).

\bibitem{kn:von66} J. von Neumann: {\em Theory of Self-reproducing Automata} (edited and completed by A. W. Burks), University of Illinois Press, Urbana and London (1966).

\bibitem{kn:Wain73} R. Wainwright (ed.): Lifeline - A Quaterly Newsletter for Enthusiasts of John Conway's Game of Life, Issues 1 to 11, March 1971 to September 1973.

\end{thebibliography}
\end{document}